\documentstyle[prl,aps,epsf,preprint]{revtex}
\tighten
\begin{document}
%\preprint{\large\bf draft}
%\draft
%\tightenlines
\title{
	Fidelity and leakage of Josephson qubits}
\author{
	Rosario Fazio$^{(1,3)}$, G. Massimo Palma$^{(2,3)}$, 
        and Jens Siewert$^{(1,3)}$}
\address{
	$^{(1)}$Dipartimento di Metodologie Fisiche e Chimiche (DMFCI), 
	Universit\`a di Catania, viale A.Doria 6, I-95125 Catania, Italy\\
	$^{(2)}$Dipartimento di Scienze Fisiche ed Astronomiche, (DSFA)
	Universit\`a di Palermo, via Archirafi 36, I-90123 Palermo, Italy\\
	$^{(3)}$Istituto Nazionale per la Fisica della Materia (INFM),
	Unit\`a di Catania e Palermo\\}

\date{\today}
\maketitle

\begin{abstract}
The unit of quantum information is  the qubit, a vector in a 
two-dimensional Hilbert space.
On the other hand, quantum hardware often operates in 
two-dimensional subspaces of vector spaces of higher dimensionality.
The presence of higher quantum states may affect
the accuracy of quantum information processing.
In this Letter we show how to cope with {\em quantum
leakage} in devices based on small Josephson junctions. 
While the presence of higher charge states of the junction 
reduces the  fidelity during gate operations 
we demonstrate that errors can be  minimized
by appropriately designing and operating the gates.
\end{abstract}


%\begin{multicols}{2}
\narrowtext
The most widely accepted paradigm of quantum computation describes 
quantum information processing in terms of quantum gates whose input 
and output are two-state quantum
systems called qubits~\cite{Deutsch95}. Quantum Computation (QC) 
is performed by means of a controllable unitary evolution of the 
qubits~\cite{Ekert96}. Due to the intrinsic quantum parallelism, 
problems which are intractable on classical computers can be
solved efficiently by using quantum algorithms. Probably the most 
striking example is the factorization of large numbers~\cite{Shor94}.

Parallel to the developement of the theory of quantum information there 
has been an increasing interest in finding physical  systems where 
quantum computation could be implemented.  In an (almost) ideal situation 
one should identify a suitable set
of two-level systems (sufficiently decoupled from any source of 
decoherence~\cite{Zurek91}) with some controllable couplings among 
them needed to realize single qubit and two-qubit operations. 
These requirements are sufficient to implement any computational
task~\cite{UniversalQC}. Various physical systems have been
suggested for the implementation of quantum algorithms, 
{\it e.g.} ions traps~\cite{Cirac95}, QED 
cavities~\cite{Turchette95} and NMR~\cite{Gershenfeld97}. 
The quest for large scale integrability and
flexibility in the design has very recently stimulated an increasing 
interest in the field of nanostructures.  Up to now  
promising proposals  are based on small-capacitance Josephson 
junctions~\cite{Shnirman97,Averin98,Makhlin99,Ioffe99,Mooij99}, 
coupled quantum dots~\cite{Loss98,Zanardi98} and phosphorus 
dopants in silicon crystals~\cite{Kane98}. The experiments 
on the superposition of charge states in Josephson 
junctions~\cite{Matters95,Bouchiat98} and the recent achievements in 
controlling the  coherent evolution of quantum states in
a Cooper pair box~\cite{Nakamura99} render superconducting nanocircuits 
interesting candidates to implement solid state quantum computers.

Physical realizations of QC are never completely decoupled from 
the environment. 
Since decoherence will ultimately limit the performance
of a quantum computer
a lot of attention is being devoted to this problem.
Besides decoherence,
for each proposed scheme a detailed analysis of the errors induced 
by the gate operations themselves is crucial in order to assess  
their reliability and the feasibility of fault-tolerant quantum 
computation~\cite{Preskill97,Kitaev97}. 
Errors may occur due to a variety of reasons.
An obvious example are fluctuations in the control 
parameters of the gate which act as a random noise 
and thus affects the unitarity of the time evolution. 
Alternatively gate 
operations can change the coupling of the qubits to the environment
(even if this coupling is negligible during storage periods)
thereby enhancing decoherence.
All these error sources can be analyzed by properly modelling the 
qubit-environment coupling. However, there are errors 
which are not due to (or cannot be described in terms of) the action of 
an external environment. Much rather they  are inherent in  the design 
of the gate.

In this Letter we consider one (intrinsic) source of error in 
gate operations which is common to several of the proposed solid state 
implementations, the {\em quantum leakage}. It 
occurs when the computational space is a subspace of a larger Hilbert 
space. This is the case {\it e.g.} when the information is encoded in 
trapped  ions or in charge (or flux) states of devices based on 
Josephson junctions (or SQUIDS). 
We start by introducing a general 
scheme to characterize the leakage and
then we focus on  devices based on small-capacitance tunnel junctions.

\begin{figure}
\vspace{3cm}
\centerline{{\epsfxsize=10cm\epsfysize=8cm\epsfbox{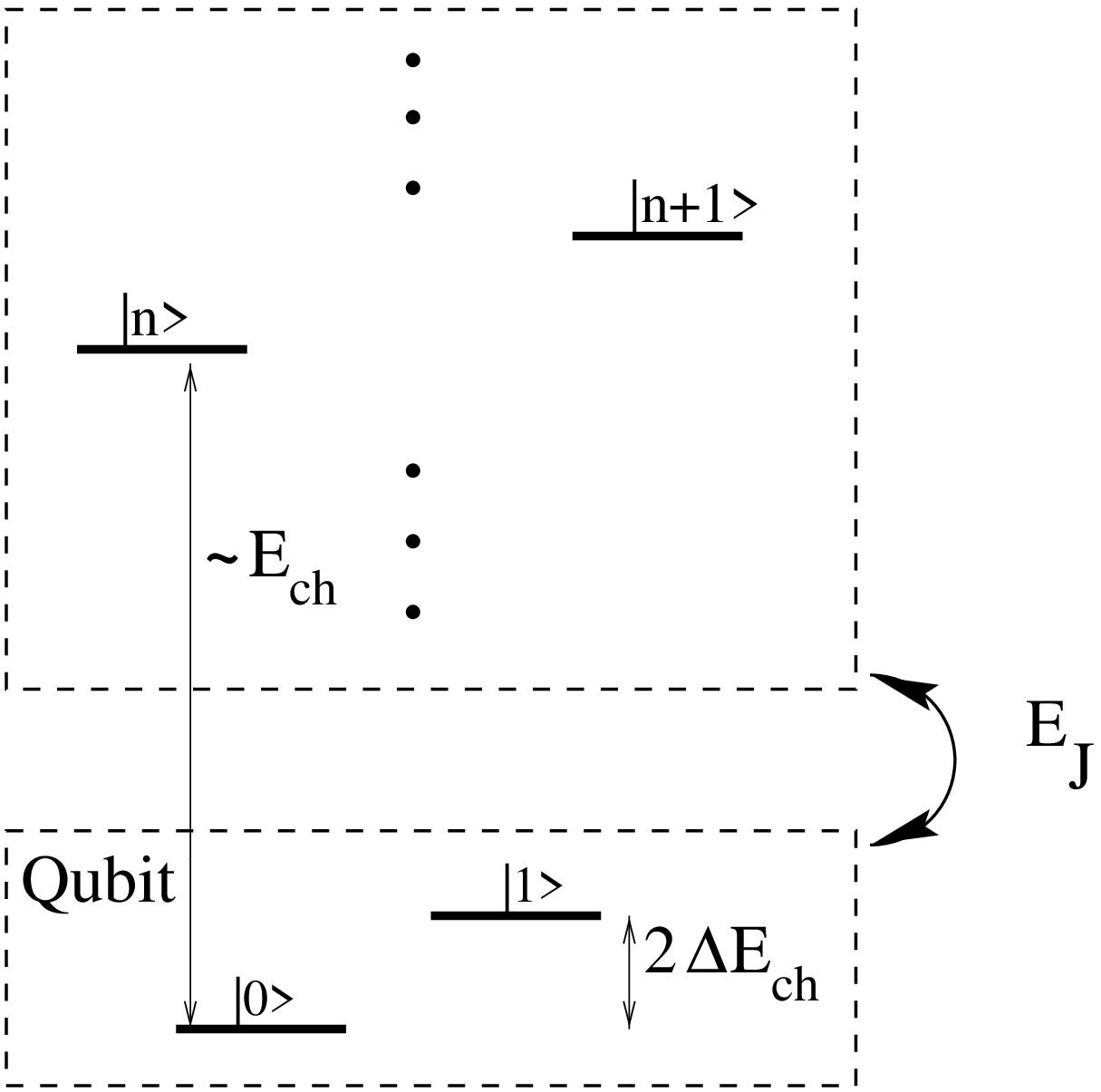}}}
\vspace{1.cm}
\caption{Schematic view of a qubit with leakage. 
         The two low energy  states
         constitute the computational Hilbert space. 
         The system however evolves with the 
         opertor $U_R$ and therefore can leak 
         out to the higher excited states. In the case 
         of Josephson junction leakage is due to the 
         Josephson tunneling to high Cooper 
         pair charge states.  In the case of two qubit operations
	 the computational space is spanned by the states 
	 $\{ |00\rangle, |01\rangle, |10\rangle, |11\rangle \}$ and the 
         coupling with the higher charge states is due both to 
         $E_J$ and to $E_L$}
\label{fig1}
\end{figure}

Our analysis applies to the situation illustrated in Fig.~\ref{fig1}. 
The two low-energy  states constitute the computational Hilbert space. 
The system, however, can leak out
to the higher states. If the energy difference between the low-lying and
the excited states is large compared to the other energy scales of the 
system (as
in Refs.~\cite{Shnirman97,Averin98,Makhlin99,Ioffe99,Mooij99}) the
probability to leak out is small. 
One might wonder whether it is necessary to discuss this effect at all.
As we will see the consequences of leakage are more severe than 
a simple estimate of energy scales might suggest.
The presence of states outside the computational space 
modifies the time evolution of the qubit states compared to 
the idealized design. 

The ideal unitary gate operation $U_I$ is obtained by switching 
on a suitable Hamiltonian $H_I$ which couples the desired 
computational states in a controlled way for a time $t_0$.
By choosing $t_0$ one can implement the desidered gate operation. 
In reality, however, the dynamics of the system 
is governed by a unitary operator $U_R$ which acts on the full Hilbert 
space. Since information is being processed within 
the computational subspace the output is related to the input state 
via the map $\Pi U_{R}(t) \Pi$, 
where $\Pi$ is the projection operator on the computational space.
One is interested in optimizing the real gate operation 
in order to get as close as possible to the ideal $U_I$. 
In general the ``best'' operation may require 
a time $t\neq t_0$ as all the system eigenenergies are modified by 
the states outside the computational subspace. 
Therefore we use the time $t$ as parameter to optimize the given computational
step. 
We characterize the performance of real gates by the 
fidelity ${\cal F}$  and the probability of leakage ${\cal L}(t)$
defined as 
\begin{equation}
	{\cal F} = 
	          1 - \frac{1}{2}\mbox{min}_{\{{\mbox t}\}} 
                      \| U_{I}(t_0) - \Pi U_{R}(t) \Pi \|  
\label{fidelity}
\end{equation}
\begin{equation} 
{\cal L}(t)  = 
                 1 -\mbox{min}_{\psi} 
                 \langle \psi |U^{\dagger}_R(t)\Pi U_R(t) 
                           |\psi\rangle
\label{leakage}
\end{equation}
In Eq.\ (\ref{fidelity}) we make use of  the operator norm defined as 
$\| D \| = \mbox{Sup}_{\psi} | D |\psi\rangle | = 
      \mbox{Sup}_{\psi} \sqrt{\langle \psi |D^{\dagger}D |\psi\rangle}$ 
over the vectors $\{ |\psi\rangle : \langle \psi|\psi\rangle =1\}$
of the computational subspace. 
This definition implies that $\| D\| = \sqrt{\lambda_M}$ where  
$\lambda_M$ is  the biggest eigenvalue of $D^{\dagger}D$. As in the case 
of the minimal fidelity~\cite{Schumacher96} this definition gives 
estimates for the worst case. 
The definition given in Eq.\ (\ref{fidelity}) can therefore be 
regarded as a prescription how 
to optimize the gate design (note that the fidelity defined 
in Eq.(\ref{fidelity}) does not depend on the time $t$).

As mentioned before the existence of states other than the 
computational ones has two main consequences on the qubit dynamics. 
There is a nonzero probability of leakage, 
measured by ${\cal L}(t)$, and a modification of the 
eigenenergies and eigenstates
of the real system. The latter effect turns 
out to be an important source of gate errors. 

In order to study the phenomena related to leakage 
quantitatively we apply Eqs.\ (\ref{fidelity}), (\ref{leakage})
to Josephson junction qubits in the charge 
regime as proposed in Refs.~\cite{Shnirman97,Makhlin99}. 
A similar analysis can be carried out, with appropriate 
changes of paramaters, for all other cases where leakage is present. 
In Refs.~\cite{Shnirman97,Makhlin99} the qubit is implemented 
using nanocircuits of Josephson junctions. The corresponding 
Hamiltonian for one and two qubit operations can be written as
\begin{equation}
	H_R  =  
	\sum_{i=1,2 } \left[  E_{\rm ch} 
	(n_{i}-n_{x,i})^2
	- E_{J}  \cos \phi_{i} \right] \nonumber \\
	 + 
	E_{L} \left( \sin \phi _1 + \sin \phi _2 \right)^2
\label{HamiltonianR}
\end{equation}
In the first term $E_{\rm ch}$ is the charging 
energy. The second and the third term represent the Josephson tunneling
(associated with the energy $E_J$) and the inductive coupling 
of strength $E_L$~\cite{footnote1}
which bring about single and two qubit operations possible.
Both $E_J$ and $E_L$ are assumed to be much smaller than
the charging energy.
The offset charge $n_{x,i}$ can be controlled 
by an external gate voltage. 
The phases $\phi_i$ and the number of Cooper pairs $n_i$
are canonically conjugate variables 
$[\phi_{i},n_{j}]= \,i \; \delta_{ij}$~\cite{footnote2}.

At temperatures much lower than the charging energy, 
for $n_{x,i} \sim 1/2$ the two
charge states $n_{i}=0,1$ are nearly degenerate. 
They represent the states $|0 \rangle $, $|1 \rangle $ 
of the qubit (see Fig.~\ref{fig1}). 
In the computational Hilbert space 
the ideal evolution of the 
system is governed by the Hamiltonian
\begin{equation}
	H_I  = 
	\sum_{i=1,2 } \left[  \Delta E_{{\rm ch},i} \sigma _{z,i}
	- \frac{E_{J}}{2} \sigma_{x,i} \right] 
	- 
	\frac{E_{L}}{2} \sigma_{y,1} \sigma_{y,2} 
\label{HamiltonianI}
\end{equation}
where $\sigma$ are Pauli matrices and 
$\Delta E_{{\rm ch},i} = E_{\rm ch}(n_{x,i} - 1/2)$. The different
time evolution due to $H_R$ and to $H_I$ causes an error
{\it in the gate operation}. We note that leakage is also present
during idle periods of the gates. However, here we only discuss
the errors during gate operations.

\underline{One-bit gate ($E_L = 0$).}
Single-qubit gate operations can be implemented, {\it e.g.},
by suddenly switching 
the offset charge to the degeneracy point $n_x =1/2$  where the charge
states $|0\rangle $ and $ |1\rangle$ are strongly
mixed by the Josephson coupling~\cite{footnote3}. 
Whereas in the ideal setup  this coupling mixes only the
states $|0\rangle$ and $|1\rangle$, in the real qubit
all charge states are involved. 

The evolution in the computational subspace
for a time interval $t$ is described by the operator ($\hbar =1$)
\begin{equation}
	\Pi U_R(t) \Pi =  \sum e^{-i E_n t } 
                           \Pi |\Phi_n \rangle \langle \Phi_n | \Pi
\end{equation}
where  $\Pi = |0\rangle \langle 0 | + |1\rangle \langle 1 | $ 
is the projector on the computational subspace and $|\Phi_n \rangle$ 
are the eigenstates with energies $E_n$ of the Hamiltonian $H_R$
(here $|\Phi_n\rangle$ can be expressed in terms of Mathieu functions). 

By evaluating the leakage according to Eq.\ (\ref{leakage}) we obtain 
\begin{eqnarray} 
	{\cal L}(t)  &=& 1 - \mbox{min}_{\pm}
	\mid \sum_{n;m=0,1} (\pm )^{m} \langle 0 \mid \Phi_n\rangle
	\langle \Phi_n \mid m\rangle e^{-iE_nt}\mid ^2 \nonumber \\
	& \sim & \frac{E_J^2}{8E_{\rm ch}^2} [1 -
	\mbox{min}_{\pm }  \cos (2 E_{\rm ch}\pm E_J/2) t ] \;\; . 
\label{onebitleakage}
\end{eqnarray}
The order of magnitude $(E_J/E_{\rm ch})^2$ can
be understood immediately by regarding the coupling to higher 
charge states as a
perturbation to the ideal system of Eq.\ (\ref{HamiltonianI}).

The fidelity has to be limited by the leakage since it
describes the length of the projection of the true state  at
time $t$ onto the ideal state at $t_0$. 
There is another effect contributing to the loss of fidelity: the 
presence of higher charge states renormalizes the energy
eigenvalues thus leading to a frequency mismatch between ideal
and real time evolution.
However, due to the symmetry of the system and the fact that
$E_J$ is the only coupling energy to the states outside the computational
subspace there is a simple way to cure this problem.
Let us consider a $\pi$-rotation.
The optimal gate is obtained by changing the operation time 
to $t_0^{\star}=\pi/\Delta E$ where $\Delta E$ is the energy splitting
between the two lowest eigenstates
(as opposed to the time $t_0 = \pi/E_J$ in the ideal system).
The value of the fidelity is then given by  
 \begin{eqnarray} 
	{\cal F} & = & 1 - \frac{1}{2}\left| 
                   \sum_{n;m=0,1} \langle\ 0 \mid \Phi_n\rangle
	\langle \Phi_n \mid m\ \rangle\ e^{-iE_nt_0^\star} -i\  
                           \right|
	\nonumber \\
	 & \sim &  1 - \frac{1}{32}\frac{E_J^2}{E_{\rm ch}^2}
	\sqrt  { 2 + 2 \mid \sin (2\pi E_{\rm ch}/E_J)\mid }
 \label{onebitfidel}
 \end{eqnarray}
We mention that the error accumulates linearly with the number of
operations.  For typical parameters of Josephson
junctions $E_J/E_{\rm ch} \sim 0.02$ one finds that
after about $10^4$ operations the loss of fidelity becomes
of order unity.

\underline{Two-bit gate ($E_L \ne 0$).}
Among the many possibilities for the elementary two-qubit operation,  
choosing a particular one may be a non-trivial step 
in the course of implementing quantum hardware.
Due to universality of quantum computation~\cite{UniversalQC}
one is free to use any generic $4\times 4$ unitary matrix
as a two-qubit gate. From our point of view a choice 
is optimal if it avoids errors stemming from a discrepancy 
of the ideal gate and the way of its implementation.
Therefore, in the following we assume that
the Hamiltonian as introduced in Eq.\ (\ref{HamiltonianI}) 
\begin{equation}
H_I = \left(     \begin{array}{cccc}
    2\Delta E_{{\rm ch}} & -E_J/2 & -E_J/2 & E_L/2  \\
    -E_J/2 &  0  & -E_L/2 & -E_J/2 \\
    -E_J/2 & -E_L/2 &  0 & -E_J/2 \\
    E_L/2  & -E_J/2 & -E_J/2 & -2\Delta E_{{\rm ch}}
                 \end{array}
      \right)\ \
\label{matHI}
\end{equation}
{\it generates} the ideal two-bit gate.
In Eq.\ (\ref{matHI}) we have used the basis 
$\{ |00\rangle, |01\rangle, |10\rangle, |11\rangle \}$
(which is obtained as the  direct product of the states introduced 
previously). The typical scale for the operation time $t_0$ is
on the order of $1/E_L$ \cite{Shnirman97,Makhlin99}.
 
In complete analogy with the one-bit gate we find that the
leakage is of the same order for the two-qubit operation: 
\[
{\cal L}\ \propto\ \max \left\{
      \left(\frac{E_J}{E_{\rm ch}}\right)^2,
      \left(\frac{E_L}{E_{\rm ch}}\right)^2 \right\}
\]
(the numerical coefficient is larger than in the one-bit case
because there are more charge states outside the computational
subspace directly coupled to the qubit states
either by $E_L$ or $E_J$).

\begin{figure}
\centerline{{\epsfxsize=12cm\epsfysize=10cm\epsfbox{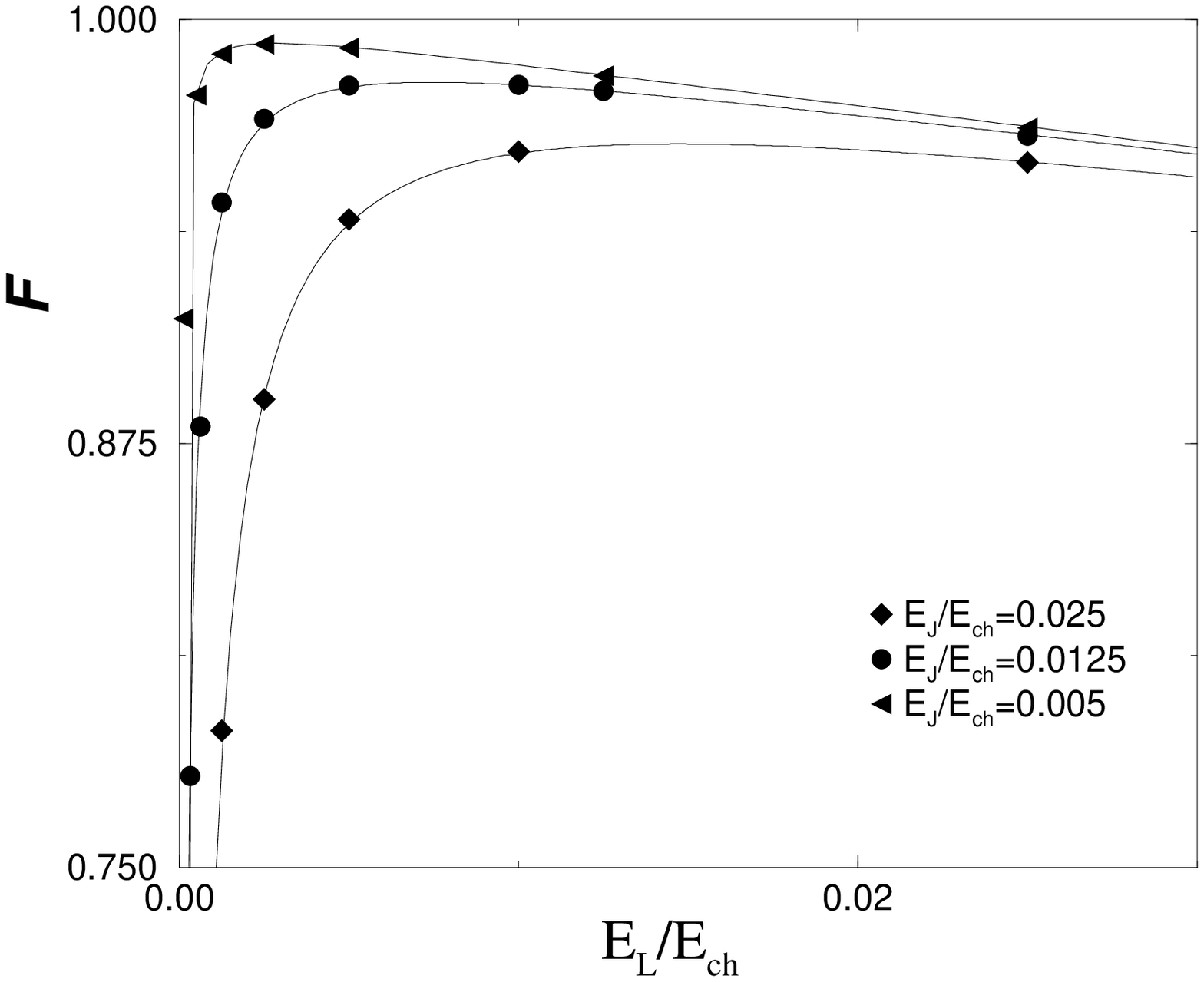}}}
\vspace{1cm}
\caption{Fidelity for a two-qubit gate as a function of $E_L$
         for different values of $E_J$.}
\label{fig2}
\end{figure}

The situation for the fidelity, however, is different. 
In order to estimate $\cal F$  we consider a perturbative
expansion of $D^{\dagger}D$ where $D=U_I(t_0)-\Pi U_R(t)\Pi$
up to second order in $E_J/E_{\rm ch}$, $E_L/E_{\rm ch}$. 
The eigenvalues of this  matrix have the
form \mbox{$2-2\cos(E_nt -E_{n}^{(0)}t_0)\ +$} 2nd order terms
(here $E_n$ and $E_{n}^{(0)}$ are the eigenvalues of $H_R$
and $H_I$, respectively). 
It turns out that 
due to  the presence of several energy scales the frequency
mismatch between real and ideal time evolution cannot be
compensated for by adjusting the operation period.
The leading terms of the fidelity can be written as
\begin{equation}
	{\cal F} \simeq 1 - \frac{1}{2}\left( a \frac{E_J^2}{E_L E_{\rm ch}} + 
                                    b \frac{E_L}{E_{\rm ch}}\right)\ \ ,
\label{twobitfidel}
\end{equation}
where $a$ and $b$ are coefficients which depend on the
particular choice of $n_{x,i}$ and $t_0$.
In Fig.~\ref{fig2} we show the numerical results 
for $n_x=1/4$ and $t_0=\pi/E_L$.  
The loss of fidelity (the term in paranthesis in
Eq.\ (\ref{twobitfidel})) is proportional to $t_0$.
The maximum (the best operation one can achieve) 
scales linearly with $E_J/E_{ch}$. 
This should be contrasted with the one-bit case 
where it scales quadratically. 

We mention that we have chosen the definitions for the leakage 
and the fidelity describing the ``worst case'' in order to avoid
a dependence of the discussion on the preparation of the initial
state. One could wonder whether the ``generic case'' is much more
robust with respect to leakage. It is easy to convince
oneself by checking various choices of initial states that
the loss of fidelity is indeed on the order of the worst
case estimates.

In conclusion, starting from given gate 
operations we have discussed their 
optimal implementation in real systems. 
We have shown that leakage limits the number
of operations which can be performed reliably both for
one and two qubit gates. For one-bit gates one can correct
leakage errors by changing the operation time. We have pointed out that
with respect to fidelity it may be appropriate to choose
the elementary two-qubit gate as it is determined by
the implementation.
Fig.~\ref{fig2} shows the central result of this work:
although leakage causes an inevitable loss of fidelity
for two-qubit operations, this loss can be minimized 
by an appropriate choice of the device parameters. 

Finally we mention that one can speculate about
correction procedures for errors caused by leakage.
It should be possible to check during the computation
whether leakage has occured. This should be done  
by measuring the  system {\em only} if it is outside the computational
subspace. One can imagine to realize a low sensitivity SET 
transistor which is able to 
measure the system only if the charge is outside a specified window.

\acknowledgments
The authors would like to thank A.K.\ Ekert, G.\ Falci, R.\ Jozsa and 
Y.\ Makhlin for helpful
discussions. This work was supported in part by the European TMR
Research Network under contracts ERB 4061PL95-1412 and FMRX-CT-97-0143.

\newpage

\newpage

%\end{multicols}

\begin{references}
\bibitem{footnote1}The coupling $E_L$ is also proportional on the 
        Josephson coupling
	energys. For our pourposes we can assume that is an independent 
        energy scale in the problem.
\bibitem{footnote2} The model ignores quasiparticle tunneling which is 
        suppressed if the temperature is much lower than the 
        BCS superconducting gap.
\bibitem{footnote3} Alternatively the qubit can be set to the degeneracy point
	adiabatically. In this paper we do not examine the various possibilities to 
        realize one and two-qbit gates. The qualitative results do not change although 
        it may be more convenient, as far as leakage is concerned, to choose a particular
	scheme.
\end{references}
\end{document}